\documentclass[12pt,paper]{JHEP3}

\usepackage{epsfig}
\usepackage{ulem}
\newcommand{\bi}{\begin{itemize}}
\newcommand{\ei}{\end{itemize}}
\newcommand{\be}{\begin{equation}}
\newcommand{\ee}{\end{equation}}
\newcommand{\bd}{\begin{displaymath}}
\newcommand{\ed}{\end{displaymath}}
\newcommand{\bea}{\begin{eqnarray}}
\newcommand{\eea}{\end{eqnarray}}
\newcommand{\g}{\gamma}
\newcommand{\ksea}{\kappa_{\mathrm{sea}}}
\newcommand{\kval}{\kappa_{\mathrm{val}}}
\newcommand{\as}{\alpha_s}
\newcommand{\nn}{\nonumber}

\newcommand{\la}{\langle}
\newcommand{\ra}{\rangle}
\newcommand{\kzb}{\overline{K^0}}
\newcommand{\kz}{K^0}
\newcommand{\bk}{B_K}
\newcommand{\bkh}{\hat{B}_K}
\newcommand{\sbar}{\overline{s}}

\newcommand{\ea}{{\em et al.}}
\newcommand{\msbar}{\overline{\mathrm{MS}}}

\newcommand{\vp}{\vec p}
\newcommand{\bkms}{\bk(\msbar,\mathrm{ 2\,GeV})}
\def\gev{\mbox{ GeV}}

\preprint{RM3-TH/04-10\\SHEP-0416}

\title{Sea quark effects in $\bk$ from $N_f=2$ clover-improved Wilson fermions}
\author{Jonathan M.~Flynn$^a$, Federico Mescia$^{b,c}$ and Abdullah Shams Bin
Tariq$^{a,d}$\\
UKQCD collaboration\vspace{0.1cm} \\
$^a$School of Physics and Astronomy, University of Southampton, \\\hspace{0.15cm}Southampton, SO17 1BJ, UK\\ 
$^b$Dipartimento di Fisica, Univ. degli Studi ``Roma TRE", \\\hspace{0.15cm}via della Vasca Navale 84, I-00146 Roma, Italy\\ 
$^c$INFN, Laboratori Nazionali di Frascati, Via E. Fermi 40, I-00044 Frascati, Italy \\
$^d$Department of Physics, Rajshahi University, Rajshahi 6205, Bangladesh\vspace{0.1cm}\\
E-mail: \email{j.flynn@hep.phys.soton.ac.uk, mescia@fis.uniroma3.it,\\asbt@hep.phys.soton.ac.uk}}

\abstract{We report calculations of the parameter $\bk$ appearing in the $\Delta S=2$ neutral kaon mixing matrix element, whose uncertainty limits the power of unitarity triangle constraints for testing the standard model or looking for new physics.  We use two flavours of dynamical clover-improved Wilson lattice fermions and look for dependence on the dynamical quark mass at fixed lattice spacing.  We see some evidence for dynamical quark effects and in particular $\bk$ decreases as the sea quark masses are reduced towards the up/down quark mass.}


\begin{document}

\section{Introduction}
$\bk$ is the $\Delta S=2$ neutral kaon mixing matrix element normalised by its vacuum saturation approximation (VSA) value,
\be
\label{eq:bkdef}
\bk(\mu) = \frac{\la\kzb|Q^{\Delta S=2}(\mu)\mid\kz\ra}{\frac{8}{3}f_K^2m_K^2},
\ee
with $\mu$ indicating the scale dependence of the operator $Q(\mu)=\sbar{\g_\mu(1-\g_5)}d\;\sbar\g^\mu(1-\g_5)d$.  This can be related to the one-loop renormalisation group invariant (RGI) value $\bkh$ through
\be
\bkh=\left[\as^{(n_f)}(\mu)\right]^{-{\frac{\g_0}{2\beta_0}}}\left[1 + \frac{\as^{(n_f)}(\mu)}{4\pi} J(n_f)\right]\bk(\mu),
\ee
where $n_f$ is the number of active flavours at the relevant scale, and $\g_0$ and $\beta_0$ have the scheme independent values of 4 and $11-2n_f/3$.  We use the $\msbar$ scheme for which $J$ is calculated to NLO in \cite{Ciuchini:1998bw}.  To go from $\msbar$ at 2 GeV to the RGI value we note that there are four active flavours and $J_{\msbar}(4)=1.792$.  Starting from the PDG value of $\Lambda_{\mathrm{QCD}}^{(5)}=216$ MeV and matching the strong coupling at the charm threshold we obtain $\bkh=1.404\,\bkms$.  This is rather insensitive to the value of $n_f$ \cite{Becirevic:2002zp}.

The standard model expression for the indirect CP violating parameter \cite{Buchalla:1996vs} as quoted in \cite{Becirevic:2000ki},
\be
\varepsilon_K = \bar{\eta}A^2\hat{B}_K\left[1.11(5)\cdot A^2(1-\bar{\rho})+0.31(5)\right],
\ee
defines a hyperbola in the $(\bar{\rho},\bar{\eta})$ plane, $A$, $\bar{\rho}$ and $\bar{\eta}$ being parameters of the CKM matrix elements and unitarity triangle \cite{Wolfenstein:1983yz}.  The theoretical uncertainty in the value of $\bkh$ remains the dominant uncertainty when we try to use this expression along with the experimental value of $\varepsilon_K$ to constrain the triangle.  This has resulted in a great deal of activity in the lattice community to refine this calculation.

\TABLE{%
\begin{tabular}{lcccccc}
\hline
\hline
 & $B_K$ & Fermion & Ren & $a^{-1}$ \\
 & $\overline{\mathrm{MS}}$, 2 GeV & Action & & (GeV) \\ 
\hline               
Kilcup \ea (1997) \cite{Kilcup:1998ye}  & 0.62(2)(2) & Staggered & Pert   & $\infty$ \\
JLQCD (1997) \cite{Aoki:1998nr}    & 0.63(4)    & Staggered & Pert   & $\infty$ \\
\hline
SPQcdR (2002) \cite{Becirevic:2002mm} & 0.66(7)   & Clover    & NP     & $\infty$ \\
JLQCD (1999) \cite{Aoki:1999gw}  & 0.69(7)   & Wilson    & NP     & $\infty$ \\
\hline
CP-PACS (2001) \cite{AliKhan:2001wr} & 0.57(1)    & DW        & Pert   & $1.8,2.8$    \\
RBC (2002) \cite{Blum:2001xb}     & 0.53(1)    & DW        & NP     & 1.9    \\
\hline
MILC (2003) \cite{DeGrand:2003in}  & 0.55(7)    & Overlap   & Pert   & $\infty$    \\
Garron \ea (2003) \cite{Garron:2003cb}   & 0.63(6)(1) & Overlap   & NP     & $2.1$    \\
\hline    
ALPHA (2003) \cite{Dimopoulos:2003kc}& 0.66(6)(2)& Tw~Mass   & NP     & $2.1$    \\
\hline
RBC (2003) \cite{Izubuchi:2003rp} & 0.50(2)    & Dyn DW    & NP     & $1.8$    \\
\hline
\hline
\end{tabular}
\caption{Some previous lattice calculations of $\bk$. NP refers to non-perturbative renormalisation.  Only the last number is unquenched.}
\label{tab:bkprev}}

There is a relatively long history of $\bk$ calculations in different frameworks.  Some recent lattice calculations are listed in table \ref{tab:bkprev}. A more comprehensive summary can be found in \cite{Gupta:2003hu}, with numbers from other methods dispersed over a relatively wide range. Over the years the quenched lattice value of $\bk$ has more or less settled down.  The 1997 quenched value of $\bkms =0.63(4)$, corresponding to $\bkh = 0.87(6)$, using staggered fermions \cite{Aoki:1998nr} remains the benchmark and is the value usually quoted for phenomenology.  Other quenched numbers are more or less consistent with this.  The error quoted however does not include any estimate for quenching effects and this has been estimated to be as high as 15$\%$ \cite{Sharpe:1997ih}.  Unquenching remains the primary systematic effect to be addressed.

There has been one preliminary report of a complete unquenched calculation using Domain Wall (DW) fermions from the RBC collaboration \cite{Izubuchi:2003rp} and a few other attempts on selected sets of parameters using Wilson and staggered fermions.  Though the central values for $\bk$ from DW fermions have often been on the lower side, the unquenched DW preliminary number is really at the lower end of the spectrum.  Other attempts to unquench, {\it e.g.}~\cite{Becirevic:2000ki,Gupta:1993bd,Kilcup:1997hp,Ishizuka:1993ya,Kilcup:1993pa,Lee:1996yg}, have not always been able to see a definite effect. However, it has been noted~\cite{Soni:1996qq} that though the unquenched numbers are consistent with the quenched numbers within errors, they are systematically lower.  Hence, it is difficult to reach an unambiguous conclusion on the true effect of dynamical fermions on this quantity. 

In this paper, we report on a calculation using two degenerate flavours of dynamical  (clover-improved) Wilson fermions.  In order to look for sea-quark dependence in $\bk$ we  use three different sea quark masses in the region $m_P/m_V\ge 0.7$ on a volume of $16^3\times 32$ ($m_{P} L\ge 7$) but a nearly constant lattice spacing. To achieve the latter a set of values of the bare coupling and bare dynamical quark mass have been chosen in ~\cite{Irving:2000hs,Allton:2001sk} to keep the lattice spacings, defined using the scale $r_0$~\cite{Sommer:1994ce}, as fixed as possible.

For $\bk$, we see some evidence for dynamical quark effects and in particular the values decrease as the sea quark mass decreases from the simulated range towards the up/down quark values.  

This calculation is undertaken as an intermediate step towards a complete unquenched evaluation of $\bk$.  In the near future one might hope to perform detailed studies over lighter and larger samples of sea quark masses at different lattice spacings in order to make the continuum extrapolation.  In the meantime, exploratory studies may help as a guide to those regions of parameters accessible today.

The plan of this paper is as follows. In section~\ref{sec:setup} we give the basic definitions and in section~\ref{sec:simul} introduce the quantities relevant for a lattice estimate of $\bk$. In section~\ref{sec:disc}, we discuss the analysis and present our values and then we have some concluding remarks in the final section.

\section{Setup of the calculation}
\label{sec:setup}

In the continuum, the operator of interest in eq.~(\ref{eq:bkdef}) is
\be
Q^{\Delta S=2}(\mu)\equiv Q_1(\mu)=\sbar\g_\mu d\;\sbar\g^\mu d+\sbar\g_\mu\g_5d\;\sbar\g^\mu\g_5d,
\ee
which is the parity conserving part of $Q(\mu)$ in eq.~(\ref{eq:bkdef}).  For Wilson fermions, owing to the breaking of explicit chiral symmetry, there is a mixing of this operator with other four-fermion operators.  Therefore one has to work with a complete basis of operators and subtract the extra ones.  One such set is 
\bea
Q_1(\mu)&=&\sbar\g_\mu d\;\sbar\g^\mu d+\sbar\g_\mu\g_5d\;\sbar\g^\mu\g_5d\nn\\
Q_2(\mu)&=&\sbar\g_\mu d\;\sbar\g^\mu d-\sbar\g_\mu\g_5d\;\sbar\g^\mu\g_5d\nn\\
Q_3(\mu)&=&\sbar d\;\sbar d+\sbar\g_5d\;\sbar\g_5d\\
Q_4(\mu)&=&\sbar d\;\sbar d-\sbar\g_5d\;\sbar\g_5d\nn\\
Q_5(\mu)&=&\sbar\sigma_{\mu\nu}d\;\sbar\sigma_{\mu\nu}d.\nn
\eea
Together with the overall multiplicative renormalisation, the subtraction of the unwanted operators may be expressed in a compact form as
\be
\label{eq:subtr}
Q^{\mathrm{cont}}(\mu) = Z (\mu, g_0^2)
\ \left( Q_1^{\mathrm{latt}} + \sum_{i\neq 1} \Delta_i(g_0^2)  Q_i^{\mathrm{latt}} \right).
\ee

The renormalisation coefficients $Z$ and $\Delta_i$ have been determined perturbatively for $\msbar$-NDR in \cite{Gupta:1997yt,Capitani:1998nj}.  Once the renormalisation and subtraction of eq.~(\ref{eq:subtr}) is carried through, we have the matrix element for our desired operator in eq.~(\ref{eq:bkdef}).

For fermion implementations which (nearly) respect chiral symmetry, {\it e.g.}~in~\cite{Kilcup:1998ye,Blum:2001xb,Garron:2003cb}, the chiral behaviour is not modified by lattice artefacts and $\bk(\mu)$ can be obtained from matrix elements of kaons at rest.  But for Wilson fermions as, for example,  in \cite{Becirevic:2000ki,Gupta:1993bd}, lattice artefacts introduce chiral symmetry breaking contributions to $\bk$ in the chiral limit. In our case, even though we use an improved-clover action, four-fermion operators are unimproved and ${\cal O}(a)$ artefacts may be present. To partially remove them at finite lattice spacing another degree of freedom is required and this can be done by introducing non-zero momentum kaons. Simulations at different lattice spacings and extrapolation to the continuum also allow lattice artefacts to be removed.

Let us now consider matrix elements with non-vanishing external momenta and generic pseudoscalar mesons. On the lattice, the chiral behaviour of the matrix element with non-vanishing external momenta can be parametrised as~\cite{Sharpe:1997ih}
\bea
\label{eq:bpchi}
\langle \bar P^0, \vec p\,\vert Q(\mu) \vert P^0, \vec q \,\rangle&=& \alpha ' +\beta ' m^2_P + \,\delta ' m_P^4 + \nonumber\\
 && (p\cdot q)\left(\gamma+\gamma ' +(\epsilon+\epsilon ')m_P^2 + (\xi+\xi ')(p\cdot q)\right) + \cdots
\eea
where all the quantities are expressed in lattice units and the ellipsis stands for higher-order terms in $ p\cdot q$ and $m_P^2$. All the primed coefficients are lattice artefacts. However, while $\g '$ and $\epsilon '$ are corrections of ${\cal O}(a)$ to the corresponding physical contributions, the parameters $\alpha '$, $\beta '$ and $\delta^\prime$ are absent in the continuum limit and have to be subtracted from the estimate of $B_K$ in eq.~(\ref{eq:bkdef}).  In particular the $\alpha '$ term makes $\bk$ divergent in the chiral limit.

For our calculation with Wilson fermions, we neglect higher order terms and use the following expression for the matrix elements:
\be
\label{eq:bpwilfit}
\langle \bar P^0, \vec p\,\vert Q(\mu) \vert P^0, \vec q\,\rangle=\alpha ' +\beta ' m^2_P + (\gamma+\gamma ')(p\cdot q).
\ee

\section{Numerical simulation}
\label{sec:simul}

In this work $\bk$ is calculated using Clover-improved Wilson fermions \cite{Luscher:1997ug} on the UKQCD set of unquenched configurations listed in table~\ref{tab:configs}. Details of the generation of the gauge configurations can be found in~\cite{Irving:2000hs,Allton:2001sk}.  To have a decorrelated sample, configurations separated by 40/50 trajectory steps are used.  These configurations are on a lattice of $32\times 16^3$ points.

\TABLE{%
\begin{tabular}{ccccccc}
\hline\hline
Set	&$\beta$& $c_{\mathrm{SW}}$& $\kappa_{\mathrm{sea}}$ & $a$(fm)[$\mbox{GeV}^{-1}$] & $(m_P/m_V)_{\ksea=\kval}$ &No.~of configs \\ 
\hline
I	&5.20	& 2.0171	& 0.1350 & 0.103(2)\,[1.91(2)]& 0.70(1)	& 100  \\ 
II	&5.26	& 1.9497	& 0.1345 & 0.104(1)\,[1.90(2)]&	0.78(1) & 100  \\ 
III	&5.29	& 1.9192	& 0.1340 & 0.102(2)\,[1.94(2)]&	0.83(1) & \phantom{0}80   \\ 
\hline\hline
\end{tabular}
\caption{The configurations used.  Values for lattice spacings are as calculated from the value of the scale, $r_0$, in lattice units from the UKQCD set \cite{Irving:2000hs,Allton:2001sk}.}
\label{tab:configs}}

The lattice spacings determined from the Sommer scale, $r_0$, are very similar for these sets.  However, there are concerns that the $\ksea$-dependence of the lattice spacing observed in these configurations is due to the proximity to a phase transition around $a\simeq 0.1$ fm where there may be large cutoff effects in the dynamical case \cite{Sommer:2003ne,DellaMorte:2004xy,Hasenbusch} and therefore needs to considered with caution.  We take the view that, nevertheless, these sets of configurations do have some degree of matching according to a valence-quark-independent definition of an effective lattice spacing, and thus, unless our physics is completely overwhelmed by any nearby phase transition, a combined analysis of the data as a function of $\ksea$ is worthwhile.  It may be noted that since $\bk$ is dimensionless, the lattice spacing enters through discretisation errors but not via an overall power of $a$. Moreover, when analysing the sea quark mass dependence, we use the variable $(am_P)^2(\ksea,\ksea)$ which in our case is equivalent to using $(r_0m_P)^2$ since our lattice spacing is defined through $r_0$ and $r_0m_P=(r_0/a)\times am_P$ with $(r_0/a)$ fixed for these lattices~\cite{Irving:2000hs,Allton:2001sk}.

Propagators and correlators were calculated using the FermiQCD \cite{DiPierro:2001yu,DiPierro:2000ve} code.  Five valence quark propagators at $\kappa =0.1356$, $0.1350$, $0.1345$, $0.1340$ and $0.1335$ were generated for each sea quark using the Stabilised Biconjugate Gradient method \cite{Frommer:1994vn}.  Smearing was tried, but since it did not give any significant improvement in the signal, the results presented here are for the local case (see comments in the next section).

To calculate the matrix element on the lattice the standard procedure \cite{Gavela:1988bd} is followed where we calculate the 3- and 2-point correlation functions
\bea
\label{eq:corr}
{\cal C}^{(3)}(t_x,t_y;p_x,p_y;\mu)&=&\sum_{\vec x\vec y}\langle\,P_5(\vec x,t_x) Q(\vec 0;\mu) P_5(\vec y,t_y)\,\rangle e^{i \vec p_x \vec x}\,e^{i \vec p_y \vec y} \\
&& \stackrel{-t_{y},t_{x}\gg 0}{\longrightarrow}{\cal Z}_P e^{- E_x t_x}\langle \overline P \vert Q(\mu) \vert P \rangle{\cal Z}_P e^{- E_y t_{y}},\nn\\\nn\\
{\cal C}^{(2)}_{J_iJ_j}(t;p_x)&=&\sum_{\vec x}\langle\,J_i(\vec x,t)J_j^\dagger(0,0)\,\rangle e^{i \vec p_x\vec x}\,\\&&\stackrel{t_x\gg 0}{\longrightarrow}\,{\cal Z}_{J_i}{\cal Z}_{J_j} e^{-E_xt}\nn.
\eea
Here the $J$'s are kaon interpolating operators and can be of pseudoscalar or axial vector type; they can also be local or smeared.  We use pseudoscalar current sources.  In the 3-pt functions, the operator is fixed at the origin and $t_y$ is kept fixed at a particular value, while $t_x$ is varied over the full temporal range of the lattice.  The main reported results are for $t_y=10$.  We have checked with other neighbouring values of $t_y$ but observe no dependence, implying that the ground state is reasonably well isolated by this time. For the momentum configurations, we have chosen $\{p_x,p_y\}=\{(0,0,0),(0,0,0)\},\{(0,0,0),(1,0,0)\}$ and $\{(1,0,0),(0,0,0)\}$ where the average over equivalent configurations is understood.  
\EPSFIGURE[t]{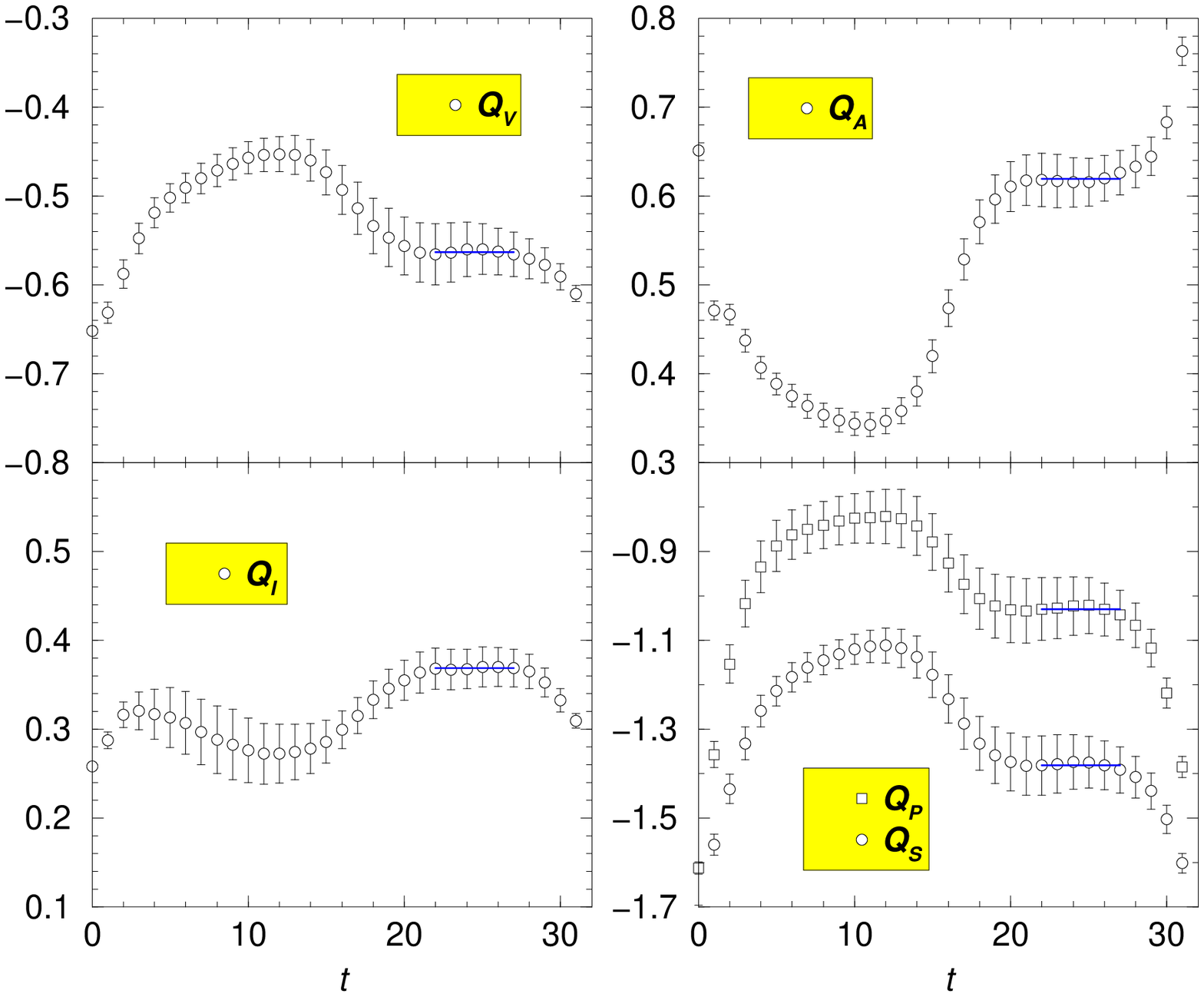,width=14cm}{Fits for lattice matrix elements for the complete set of bare operators for a sample of our data (set I, $\kval=0.1350$).  Ratios of the 3-pt correlators to two 2-pt $\langle PP\rangle$ correlators are fitted in the interval $t_x=22-27$ for $t_y=10$ (see eq.~\ref{eq:ratio}). Correlators are shown for zero momentum.  The fitted ones are those of interest $\langle \bar P^0|Q_i|P^0\rangle$ while the other plateau in the first half of the lattice corresponds to the $\langle \bar P^0 \bar P^0|Q_i|0\rangle$ matrix elements.
\label{fig:ratiofit}}

For simulation, we use the simpler basis of 
\bea
Q_V(\mu)&=&\sbar\g_\mu d\;\sbar\g^\mu d\nn\\
Q_A(\mu)&=&\sbar\g_\mu\g_5d\;\sbar\g^\mu\g_5d\nn\\
Q_I(\mu)&=&\sbar d\;\sbar d\\
Q_P(\mu)&=&\sbar\g_5d\;\sbar\g_5d\nn\\
Q_S(\mu)&=&\sbar\sigma_{\mu\nu}d\;\sbar\sigma_{\mu\nu}d,\nn
\eea
which is related to our renormalisation basis introduced in the previous section through a simple rotation.  Fitted ratios for this basis that give us the matrix elements in lattice units, $Q_i^{\mathrm{latt}},\;(i=V,A,I,P,S)$ are plotted in fig.~\ref{fig:ratiofit}. 

To go directly to $\msbar$ at $\mu = 2$ GeV, we note that in our case $(a\mu)\approx 1$ and we can naively use standard perturbation theory at one-loop.  For the coupling there is a range of choices that may lead to different numerical values.  We use the boosted bare lattice coupling, $g_0^2=6/\beta\langle P\rangle$, where $\langle P\rangle$ is value of the relevant average plaquette and our values are $\{0.5336,$ $0.5399,\,0.5424\}$. For $c_{SW}$ the one-loop value of 1.0 is used.  The perturbative matching coefficients thus obtained are listed in table \ref{tab:ren}.

\TABLE[t]{%
\begin{tabular}{cccccccc}
\hline\hline
Set &$g_0^2$ &$Z(2\gev,g_0^2)$ &$Z\,\Delta_1(g_0^2)$ &$Z\,\Delta_2(g_0^2)$ &$Z\,\Delta_3(g_0^2)$& $Z\,\Delta_4(g_0^2)$ &$Z_A$\\
\hline
I   &2.162   &0.4959 &$-0.0385$ &$-0.0070$ &0.0140 &0.0140 &0.7482\\
II  &2.113   &0.5072 &$-0.0376$ &$-0.0068$ &0.0137 &0.0137 &0.7540\\
III &2.091   &0.5133 &$-0.0372$ &$-0.0068$ &0.0135 &0.0135 &0.7565\\
\hline\hline
\end{tabular}
\caption{Perturbative matching coefficients to go from $\bk^{\mathrm{latt}}(\mu=1/a)$ to $\bk^{\msbar}(\mu=2$ GeV).}
\label{tab:ren}}

To extract the desired matrix element the following ratios are formed:
\bea
\label{eq:ratio}
R_3=\frac{{\cal C}^{(3)}(t_x,t_y;p_x,p_y;\mu)}{Z^2_A{\cal C}^{(2)}_{PP}(t_x;p_x){\cal C}^{(2)}_{PP}(t_y;p_y)}&\longrightarrow&\frac{1}{Z_A^2{\cal Z}_P^2}\,\langle \bar P^0, \vec p_x \vert Q(\mu) \vert P^0, \vec p_y\rangle ,\\\nn\\
X(0)=\frac{8}{3}\left|\frac{{\cal C}^{(2)}_{A_0P}(t_x)}{{\cal C}^{(2)}_{PP}(t_x)}\right|^2&\longrightarrow&\frac{1}{Z_A^2{\cal Z}_P^2}\frac{8}{3} f_P^2 m_P^2,\\\nn\\
X(\vp)=X(0)\cdot\frac{(p_x\cdot p_y)}{m_P^2}&\longrightarrow&\frac{1}{Z_A^2{\cal Z}_P^2}\frac{8}{3}f_P^2(p_x\cdot p_y),
\eea
where $Z_A$ is the axial current renormalisation.

At this stage one may fit the equation
\be
\label{eq:romefit}
R_3=\widetilde\alpha '+\widetilde\beta 'X(0)+(\widetilde\g + \widetilde\g ') X(\vp),
\ee
with
\bd
\widetilde\alpha ' \equiv \frac{\alpha '}{Z_A^2{\cal Z}_P^2},\;\;\widetilde\beta '   \equiv \frac{3\beta '}{8f_P^2},\;\;\widetilde\g \equiv \frac{3\g}{8f_P^2}\;\;\mbox{and}\;\;\widetilde\g ' \equiv\frac{3\g '}{8f_P^2},
\ed
to obtain estimates for $\bk$ from $\widetilde\g$~\cite{Becirevic:2002mm,Crisafulli:1996ad}, by neglecting $\widetilde\g '$.  In the fit, the parameters with tildes are taken to be constant and hence the estimates are for effective values of $\cal{Z}_P$ and $f_P$ in our range of simulation.  In this manner, for a set of different valence quarks with a given sea quark mass, this approach gives an estimate of the leading term in an expansion of $\bk$ for that set with the kaons not necessarily being at the physical kaon mass.

To obtain estimates of $\bk$ for each $(\ksea,\kval)$ combination, which will then allow us to extrapolate in the quark masses, we follow the approach of \cite{Gupta:1993bd,Gupta:1997yt}.  Let us call the non-zero- and zero-momentum $R_3$'s $R_3(\vp)$ and $R_3(0)$ respectively, corresponding to $X(\vp)$ and $X(0)$ defined in eq.~\ref{eq:ratio}. The two non-zero momenta $\{p_x,p_y\}=\{(0,0,0),(1,0,0)\}$ and $\{(1,0,0),(0,0,0)\}$, have been averaged, since they are estimates of the same matrix elements in the continuum and indeed numerically are found to be very similar.  Then we have 
\be
\label{eq:bkkskv}
\left.\frac{R_3(\vp)-R_3(0)}{X(\vp)-X(0)}\right|_{(\ksea,\kval)}=\bk(\mu,\ksea,\kval).
\ee
These can then be used in our chiral extrapolations in the sea and valence quarks.  At the same time, fitting these values to a constant for a given sea quark is similar to estimating $\widetilde\g$ from a fit of eq.~\ref{eq:romefit}.  At higher orders of momentum, this expression differs from the correct dependence of $\bk$ by a term like $\widetilde\xi m_PE(\vec{p})$~\cite{Gupta:1997yt}.  We have found the coefficient $\widetilde\xi$ of this term difficult to determine, particularly for our limited set of momenta.  However, if we were able to make this correction, it would simply change our values of $\bk$ within our systematics, leaving our conclusions unchanged.

\section{Analysis and discussion}
\label{sec:disc}

The values obtained for $\bk(\msbar, 2$ GeV) for our sets of masses  are tabulated in table~\ref{tab:bkkskv}. We refer to the ones quoted from eq.~(\ref{eq:romefit}) following~\cite{Becirevic:2002mm,Crisafulli:1996ad} and from eq.~(\ref{eq:bkkskv}) following~\cite{Gupta:1993bd,Gupta:1997yt} as method I and II respectively.

We have degenerate valence quarks.  So, SU(3) breaking effects due to $m_s\neq m_{u,d}$ are not accounted for.  Rather our kaon is made up of two quarks around $m_s/2$. Moreover, the results in table~\ref{tab:bkkskv} are obtained for local sources. Indeed, we have not seen any significant improvement of the signal from smearing.  This is not unexpected since we have a local operator at the origin and can smear only at the sink, which is usually less effective than source smearing.  It may also be due to a lack of optimisation of the smearing parameters.  However, results were fully compatible with those using local operators and we have restricted the presentation to the simpler case.

In fig.~\ref{fig:bkkskv}, we plot $\bk(\msbar, 2$ GeV$)$ from method II as a function of the corresponding squared pseudoscalar masses over the complete set of our valence and sea quark masses.  We observe the points for the lightest valence quarks diverging for the different sea quarks. Here, it may be noted that for $\kval\gg\ksea$ the theory becomes more like quenched.  This effect is clearly seen in fig.~3 of \cite{Becirevic:2003wk} where as the valence quark becomes lighter the partially quenched curves leave the full theory and tend towards the quenched one. Finite volume effects are, in general, expected to be small~\cite{Sharpe:1992ft}, but for some regions of parameter space, particularly for very light quarks, it has been suggested that finite volume effects can obscure the chiral behaviour in $\bk$ \cite{Becirevic:2003wk}.  The JLQCD collaboration~\cite{Aoki:2002uc} observes finite volume effects for lighter sea quarks for the same action, but for our parameters they have excluded finite volume effects for pseudoscalar meson correlators down to just beyond our lightest point in set I.  Indeed we find the finite volume correction from \cite{Becirevic:2003wk} to be $-0.1\%$ for this point.  Nonetheless, we note that, contrary to the other sets, for set I, the ${\cal O}(a)$ discretisation error parameters $\widetilde\alpha$ and $\widetilde\beta$ turn out to have finite values of $-0.06(2)$ and 0.23(8).  The effects of these terms are greater at lighter quark masses and we cannot be sure that the curvature observed here is due to a true chiral behaviour. As can be seen from our values of $m_P/m_V$, this point is at a considerably lighter mass than all the other points.  Therefore, we choose to be cautious and exclude it from our analysis. It would be interesting to know if non-perturbative renormalisation~\cite{Martinelli:1995ty,Donini:1999sf}, and/or the improvement programme of~\cite{Frezzotti:2003ni,Frezzotti:2003zc} could lead to better chiral behaviour.

\TABLE[t]{%
\begin{tabular}{cccccc}
\hline\hline
&&&&Method I &Method II\\\cline{5-5}\cline{6-6}
$(\beta, \ksea)$&$\kval$&$m_P/m_V$ & $(am_P)^2$&$\bk(\ksea)$&$\bk(\ksea,\kval)$  \\ 
\hline
(5.20, 0.1350) 	 & 0.1356 &0.62(3) &0.106(5)&0.64(7)& 0.41(12) \\
		 & 0.1350 &0.72(2) &0.166(4)&       & 0.57(9)\phantom{1}  \\
		 & 0.1345 &0.77(1) &0.218(4)&       & 0.63(7)\phantom{1}  \\
		 & 0.1340 &0.80(1) &0.270(4)&       & 0.66(6)\phantom{1}  \\
		 & 0.1335 &0.83(1) &0.324(4)&       & 0.69(5)\phantom{1}  \\
\hline
(5.26, 0.1345) 	 & 0.1356 &0.67(2) &0.151(3)&0.69(8)& 0.70(16) \\
		 & 0.1350 &0.74(1) &0.206(3)&       & 0.71(10) \\
		 & 0.1345 &0.77(1) &0.255(3)&       & 0.71(8)\phantom{1}  \\
		 & 0.1340 &0.81(1) &0.306(4)&       & 0.72(7)\phantom{1}  \\
		 & 0.1335 &0.83(1) &0.359(4)&       & 0.72(6)\phantom{1}  \\
\hline
(5.29, 0.1340) 	 & 0.1356 &0.72(2) &0.170(5)&0.79(4)& 0.81(6)\phantom{1}  \\
		 & 0.1350 &0.77(1) &0.229(5)&       & 0.79(4)\phantom{1}  \\
		 & 0.1345 &0.80(1) &0.280(5)&       & 0.78(4)\phantom{1}  \\
		 & 0.1340 &0.83(1) &0.332(6)&       & 0.77(4)\phantom{1}  \\
		 & 0.1335 &0.85(1) &0.386(6)&       & 0.77(4)\phantom{1}  \\
\hline\hline
\end{tabular}
\caption{Simulated values of $\bk(\msbar, 2$ GeV). Method I refers to a direct fit of eq.~(\ref{eq:romefit}); while in method II, eq.~({\ref{eq:bkkskv}}) is used to obtain values for each $(\ksea,\kval)$ combination.}
\label{tab:bkkskv}}

\EPSFIGURE[ht]{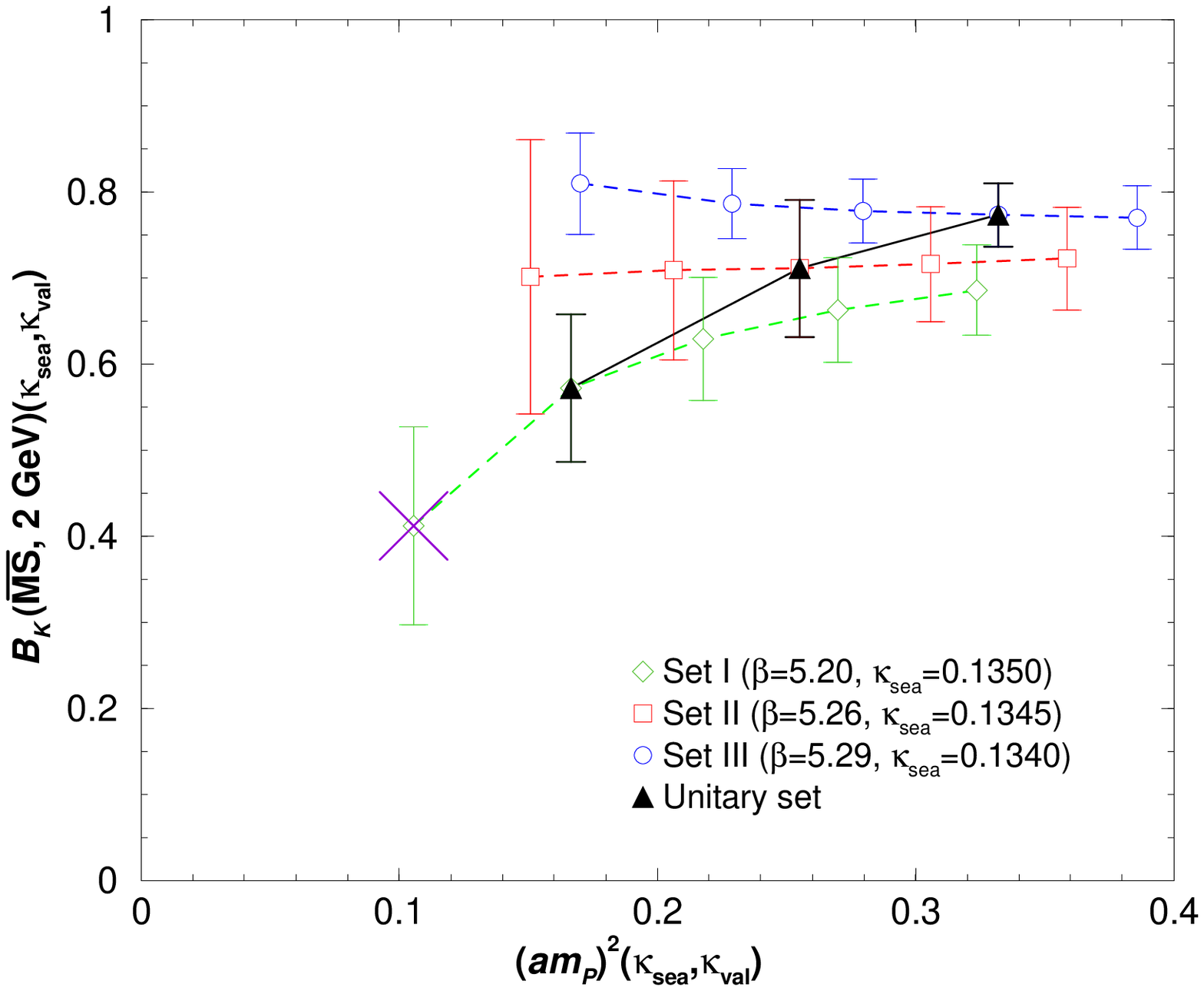,width=14cm}{Values of $\bk(\msbar, 2$ GeV) for each $(\ksea,\kval)$ combination plotted as a function of the corresponding squared pseudoscalar masses.  The dashed lines joining the points are just for a visual guide separating the sets with different sea quarks.  The filled points joined by a solid line are the unitary ones for which $\ksea=\kval$.  The lightest point for set I (marked by a large cross) is excluded from the analysis.\label{fig:bkkskv}}

Now let us consider the values from method I. It is notable that for these rather heavy sea quarks these numbers are compatible with quenched estimates.  This is the reason that previous attempts to unquench for a fixed heavy sea quark mass have found it difficult to disentangle the unquenching effects.

Since we have more than one sea quark mass in our simulation, we can attempt to extrapolate these numbers to the chiral limit. We use a linear fit versus the unitary pseudoscalar masses $(am_P)^2(\ksea=\kval)$ and go to the up/down limit.  This gives us
\be
\label{eq:bkmethodI}
\bk\mathrm{(\msbar,\;2\;GeV)} = 0.49(13),
\ee
which corresponds to $\bkh = 0.69(18)$.
\EPSFIGURE[ht]{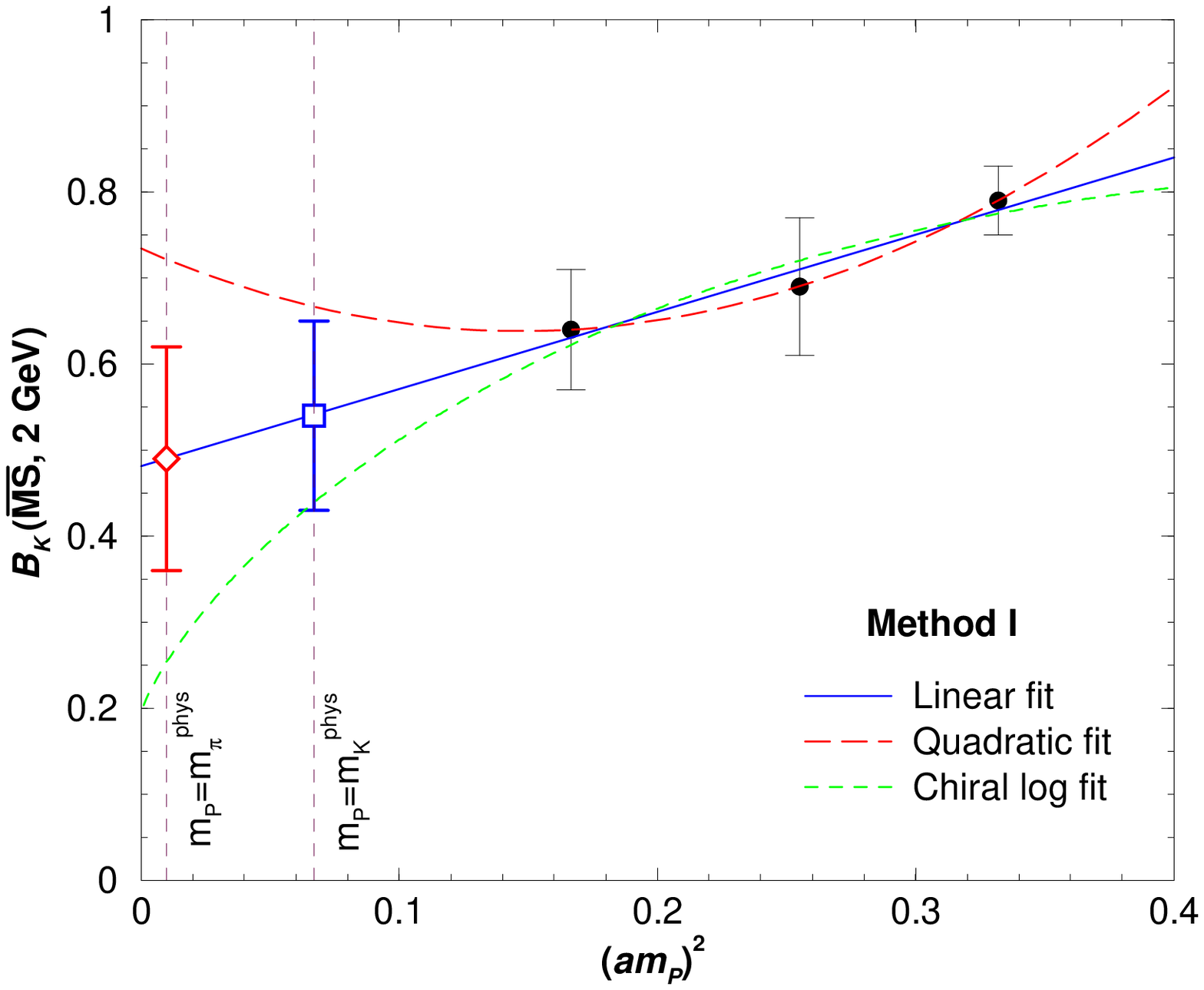,width=14cm}{Fit to the data from method I.  The values quoted is from the linear extrapolation, whereas the quadratic and chiral log-type fits are added for illustration.  The extrapolated points at $m_P=m_\pi^{\mathrm{phys}}$ and $m_P=m_K^{\mathrm{phys}}$ are also shown.}

In this method we estimate $\widetilde\g$ in eq.~\ref{eq:romefit}.  As mentioned in the previous section, the valence quarks are not necessarily such that $m_P=m_K^{\mathrm{phys}}$.  In fact one can note by comparing with the last column of table~\ref{tab:bkkskv} that these values are in the simulated region.  Therefore one may think of this estimate as one of $\bk$ where the sea quarks are realistically light but the valence quarks are heavier than the physical strange quark.

A somewhat complementary approach, would be to follow the route of \cite{Izubuchi:2003rp} and take the unitary points, {\it i.e.}~the points with $\ksea=\kval$ from method II, for extrapolation to the physical kaon mass [fig.~\ref{fig:bkuni}].  This leads to
\be
\label{eq:bkmethodII}
\bk\mathrm{(\msbar,\;2\;GeV)} = 0.48(13),
\ee
Corresponding to $\bkh = 0.67(18)$.
\EPSFIGURE[ht]{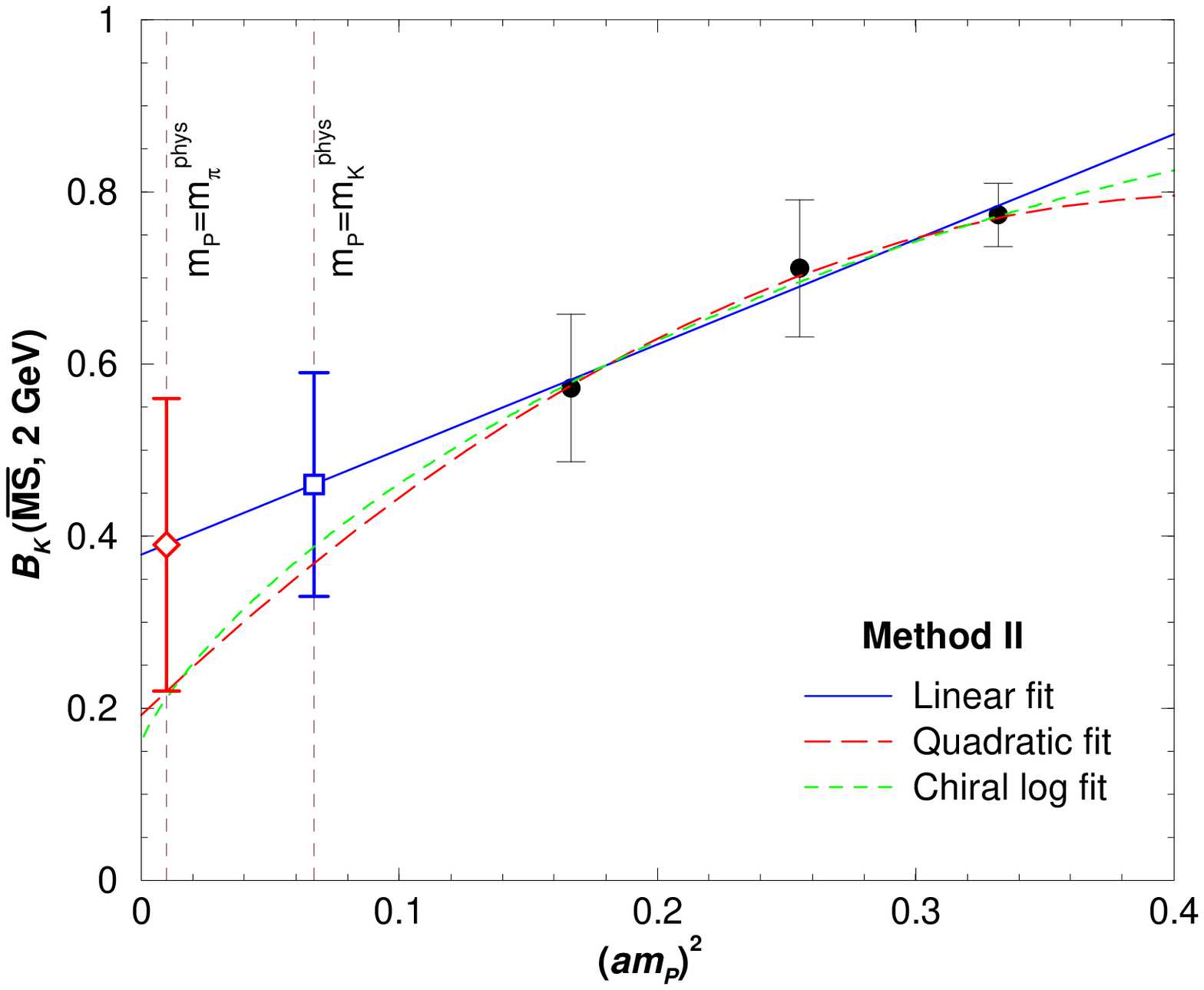,width=14cm}{Unitary fit of the data.  The value quoted is from the linear extrapolation, whereas the quadratic and chiral log-type fits are added for illustration.  The vertical lines show the positions where $m_P=m_\pi^{\mathrm{phys}}$ and $m_P=m_K^{\mathrm{phys}}$.\label{fig:bkuni} respectively.}
Here we have a more reasonable valence $m_P=m_K^{\mathrm{phys}}$, but on the other hand the sea and valence quarks are degenerate and hence the sea content is not as light as the up/down quarks. To understand how much this may affect us we note that if we take all the quark masses (both valence and sea) to zero our value of $\bk$ goes down to 
$0.40(17)$ and $\bkh = 0.56(24)$.

A combined analysis of valence and sea quarks has been tried for the spectroscopy studies in~\cite{Allton:2001sk,Aoki:2002uc}.  With more momenta, higher statistics and/or a larger sample of sea and valence quark masses and if the higher order terms in $\bk$ could be estimated, this would be a possible route to an estimate of $\bk$ at the physical valence and sea masses. 

Even though we recognise that the presence of several artefacts does not allow a quantitative estimate of the sea quark dependence, it does seem that dynamical quark effects can be quite significant. There also seem to be indications that, incorporating dynamical quarks lowers the value of $\bk$. Taking this together with the observation in~\cite{Soni:1996qq} that the $N_f=2$ numbers are always lower than those for $N_f=0$, a statement also valid for subsequent works, we see that when one has two finite-mass but still heavy sea quarks, $\bk$ starts to decrease but is still consistent with the quenched value within errors.  When the sea quarks can be taken to the massless limit, the value of $\bk$ becomes distinctly lower than the quenched result. It is also intriguing to note that in a recent study where $\bkh$ is taken as a free parameter and fitted using the other unitarity triangle constraints, the value obtained is $\bkh=0.69(11)$~\cite{Ciuchini:2003rk}, again lower than the usual quenched lattice value.

Owing to the exploratory nature of our analysis and large statistical errors, a study of systematic errors such as those connected to choices of fit window, chiral extrapolation, renormalisation method, the fixed time at one end, the strong coupling, $\Lambda_{QCD}$, etc. has not been addressed.

\section{Conclusion}
We have presented results for $\bk$ calculated using
non-perturbatively $\mathcal{O}(a)$-improved Wilson fermions with two
dynamical flavours for three sets of relatively small volume lattices
of matched spacing. Despite some concern about the robustness of the
estimates due to various lattice uncertainties, there are indications
that dynamical quark effects are important and lead to a lower value
of $\bk$.

\acknowledgments We would like to thank Massimo Di Pierro for his help
in using the FermiQCD code. We also thank the Iridis parallel
computing team at University of Southampton, in particular, Ivan
Wolton, Ian Hardy and Oz Parchment for their computing support. We
thank Damir Becirevic, Ken Bowler, Martin Hasenbusch, Alan Irving,
Laurent Lellouch, David Lin, Vittorio Lubicz, Craig McNeile, Chris
Michael, Amarjit Soni and Giovanni Villadoro for their comments. The work of ASBT is
supported by a Commonwealth Scholarship.  Work partially supported by
the European Community's Human Potential Programme under
HPRN-CT-2000-00145 Hadrons/Lattice QCD.  FM is also partially supported by IHP-RTN, EC contract no. HPRN-CT-2002-00311
(EURIDICE).

\bibliographystyle{JHEP}
\bibliography{bk}

\end{document}